\documentclass[twocolumn]{revtex4}
\usepackage[T1]{fontenc}
\usepackage{graphicx}
\usepackage{rotating}
\usepackage{bm}        
\usepackage{amssymb}   
\usepackage{bm}
\usepackage{float}
\usepackage{tikz}
\usepackage{amsmath}

\def\jcp#1#2#3{{ J.~Chem.~Phys.}~{\bf #1},\ #2\ (#3)}

\def\prl#1#2#3{{ Phys.~Rev.~Lett.}~{\bf #1},\ #2\ (#3)}

\begin{document}
\include{epsf}


\title{Machine-learning-corrected quantum dynamics calculations}

\author{A. Jasinski}
\affiliation{Department of Physics, Penn State University,
Berks Campus, Reading, PA 19610-6009}

\author{J. Montaner}
\affiliation{Department of Physics, Penn State University,
Berks Campus, Reading, PA 19610-6009}

\author{R. C. Forrey}
\email{rcf6@psu.edu}
\affiliation{Department of Physics, Penn State University,
Berks Campus, Reading, PA 19610-6009}

\author{B. H. Yang}
\affiliation{Department of Physics and Astronomy and the Center for
Simulational Physics, University of Georgia, Athens, Georgia 30602}

\author{P. C. Stancil}
\affiliation{Department of Physics and Astronomy and the Center for
Simulational Physics, University of Georgia, Athens, Georgia 30602}

\author{N. Balakrishnan}
\affiliation{Department of Chemistry and Biochemistry, University of Nevada,
Las Vegas, NV 89154}

\author{J. Dai}
\affiliation{Department of Chemistry, University of British Columbia,
Vancouver, B. C. V6T 1Z1, Canada}

\author{R. A. Vargas-Hern\'{a}ndez}
\affiliation{Department of Chemistry, University of British Columbia,
Vancouver, B. C. V6T 1Z1, Canada}

\author{R. V. Krems}
\email{rkrems@chem.ubc.ca}
\affiliation{Department of Chemistry, University of British Columbia,
Vancouver, B. C. V6T 1Z1, Canada}

\date{\today}

\begin{abstract}
Quantum scattering calculations for all but low-dimensional systems at low energies must rely on approximations. All approximations introduce errors. The impact of these errors is often difficult to assess because they depend on the Hamiltonian parameters and the particular observable under study. Here, we illustrate a general, system and approximation-independent, approach to improve the accuracy of quantum dynamics approximations. The method is based on a Bayesian machine learning (BML) algorithm that is trained by a small number of rigorous results and a large number of approximate calculations, resulting in ML models that accurately capture the dependence of the dynamics results on the quantum dynamics parameters. Most importantly, the present work demonstrates that the BML models can generalize quantum results to different dynamical processes. Thus, a ML model trained by a combination of approximate and rigorous results for a certain inelastic transition can make accurate predictions for different transitions without rigorous calculations. This opens the possibility of improving the accuracy of approximate calculations for quantum transitions that are out of reach of rigorous scattering calculations.

\end{abstract}


\maketitle

\section{Introduction}

Quantum dynamics problems with time-independent Hamiltonians are often solved by representing the Hamiltonian eigenstates by a basis set expansion and numerically integrating the resulting set of coupled equations. For example, in molecular scattering theory, the scattering matrices are computed by integrating coupled channel equations \cite{alex}; single-particle and few-particle Green's functions in lattice systems can be computed by solving coupled recursive equations \cite{mona}; the bound states of few-body quantum systems can be computed by matching the solutions to coupled differential equations with different initial conditions \cite{jeremy}. As the complexity of quantum systems increases, the number of coupled equations required to obtain accurate solutions becomes prohibitively large. Therefore, quantum dynamics calculations are often based on decoupling approximations that reduce the problem to smaller, independent sets of coupled equations. These approximations necessarily introduce errors into the dynamical results. The impact of these errors is often difficult to assess because they depend on the Hamiltonian parameters and the particular observable under study. In this work, we demonstrate a general, system and approximation-independent, method to enhance the accuracy of the decoupling approximations. The approach is based on Bayesian model calibration \cite{bmc}, originally introduced to calibrate computer simulation models by experimental observations. Here, we show that a similar method can be used to build Bayesian machine learning (BML) models that learn correlations between the approximate and rigorous results for some dynamical processes and transfer this information to correct the quantum dynamics approximations for other processes.


We consider two different problems: (i) the inelastic scattering problem of two diatomic molecules prepared in a wide range of internal quantum states and undergoing collisions in a wide range of collision energies; (ii) atom - diatom chemically reactive scattering in a wide range of energies and angular momentum states.   
Accurate predictions of probabilities for such collisions are required for applications in astrophysics \cite{lique}, planetary atmosphere models, the development of new crossed-beam experiments exploiting control over the longitudinal motion of molecular beams for precision measurements \cite{Perreault2017}, cold chemistry \cite{Krems2008, Bala2016} as well as the mechanistic understanding of microscopic collision dynamics \cite{Jambrina2019, my-book}. Rigorous quantum calculations of inelastic molecule - molecule scattering must be performed in six nuclear dimensions and account for all  couplings between the internal and translational motion states of the colliding molecules. The complexity of the problem can be reduced by eliminating some of the angular momentum couplings and/or freezing some degrees of freedom. We consider an approximation that reduces the active configuration space to five dimensions (5D) and neglects angular momentum couplings giving rise to Coriolis interactions. For the reactive scattering problem, we consider a popular $J$-shifting approximation \cite{Bowman1991, Zhang1999}, which produces reaction cross sections for high total angular momentum ($J$) scattering based on rigorous results  for $J=0$. In both cases, 
we compare the performance of the approximate dynamical approaches with the BML-corrected results, demonstrating that BML can be used to enhance the accuracy of approximate quantum dynamics methods.


Machine learning has been previously combined with quantum calculations in order to solve problems in quantum condensed-matter physics \cite{qcp-1,qcp-2, qcp-3}, quantum chemistry \cite{qc-1,qc-2,qc-3,qc-4,qc-5,qc-6,qc-7} and molecular dynamics \cite{md-1, md-2}. These studies can be classified into approaches based on artificial neutral networks (NN) and kernel-based methods, including Bayesian ML using Gaussian processes (GP)  \cite{gp-book}. NNs generally require a large number of input data to produce accurate models. In the present work, our goal is to build models of quantum dynamical observables improved by a small number of rigorous calculations. Previous applications show that Bayesian ML based on GPs can produce powerful, non-parametric prediction models based on very sparse data \cite{extrapolation-1,extrapolation-2,extrapolation-3, NNs-vs-GPs, rodrigo-bo, bml}. This makes GPs ideally well suited for Bayesian model calibration (BMC) \cite{bmc, bml}, which aims to compensate for the (unknown) deficiencies  of a simulation model in a flexible, non-parametric way. In the context of quantum dynamics, BMC has been used to interpolate quantum results by a ML model trained with a large number of classical trajectory calculations \cite{krems2}. BMC has also been used in quantum chemistry to enhance the accuracy of potential energy calculations \cite{qc-bmc}. The present work employs BMC to improve quantum dynamics approximations. 


First, we consider state-resolved inelastic collisions between two diatomic molecules in well-defined quantum states specified by the vibrational ($v$) and rotational ($j$) quantum numbers
\begin{eqnarray}
A(v_1,j_1) + B(v_2,j_2) \rightarrow A(v'_1,j'_1) + B(v'_2,j'_2), 
\end{eqnarray}
where the subscripts are used to label the different molecules. 
Our goal is to build a ML model of cross sections for such collisions.   
We impose the following requirements on the model: the model must be easy to evaluate; the model must be more accurate than the results of the  approximate dynamical calculations; the model must be non-parametric to adapt to increasing information about the scattering process, which may reveal new dynamical features, such as resonances; the model must require as little information from rigorous quantum calculations as possible.    

Our ML models are based on GPs so we begin by a brief description of a GP (see Refs. \cite{gp-book, bml} for more details). 
The purpose of a GP model is to make a prediction of some quantity $y$ at an arbitrary point $\bm x $ of a $\cal D$-dimensional space, given a finite number $n$ of values $\bm y = \left ( y_1, ..., y_n \right )^\top$, where $y_i$ is  the value of $y$ at $\bm x_i$.  In the absence of noise in $y_i$, the goal is to infer the function $f(\bm x)$ that interpolates $y_i \Leftarrow f(x_i)$.  
GPs  infer the conditional {distribution over functions} $p(f| {\bm y})$. 
The conditional mean of  such distribution as a function of $\bm x_\ast$ is given by  \cite{gp-book}
\begin{eqnarray}
\mu(\bm x_*) &=& K({\bm x_*},\bm x)^\top \left [ K(\bm x, \bm x) + \sigma_n^2 I \right ] ^{-1}{\bm y} \label{eqn:gp_mu}
\label{meanGP}
\end{eqnarray} 
where  $\bm x_\ast$ is a point in the input space where the prediction is to be made, 
 $K(\bm x, \bm x)$ is the $n \times n$ square matrix with the elements $K_{i,j} = k(\bm x_i,\bm x_j)$. The function $k(\bm x',\bm x'')$ represents the covariance between the normal distributions of $y$ at $\bm x'$ and $\bm x''$. 
The unknown parameters of this function are found 
by maximizing the log \emph{marginal likelihood} function,
\begin{eqnarray}
\log p(\bm{y}|\bm{\theta}) = -\frac{1}{2} \bm{y}^\top K^{-1} \bm{y} - \frac{1}{2}\log |K| -\frac{n}{2} \log (2\pi),
\label{eqn:logml}
\end{eqnarray}
where $\bm \theta$ denotes collectively the parameters of $k$ and $|K|$ is the determinant of the matrix $K$. Given the $k$ functions thus found, Eq. (\ref{meanGP}) is a GP model. 

Following Refs. \cite{bmc,krems2}, we now define the ML model of a quantum dynamics cross section as 
\begin{equation}
\sigma_{\rm ML} (\bm x) =a\,{\cal F}( \bm x )+{\cal G}(\bm x)
\label{model}
\end{equation}
where $\cal F$ and $\cal G$ are independent GPs,
$a$ is a variable parameter, and $\bm x$ is a vector of input variables.
Here $\cal F$ is designed to describe the $\bm x$-dependence of approximate dynamical results and
${\cal G}(\bm x)$ infers the difference between the
 approximate and accurate calculations. We will compare the accuracy of $\sigma_{\rm ML}$ thus obtained with the cross sections $\sigma$ obtained directly from approximate calculations.

To build a general ML model $\sigma_{ML}$, we define the input variable space $\bm x$ as follows: 
\begin{equation}
\bm x=\{E_c,\,\Delta E_{int},\,\Delta A_{int},\,\Delta v_1,\,\Delta j_1,
\,\Delta v_2,\,\Delta j_2\}
\label{xset}
\end{equation}
where $E_c$ is the collision energy, $\Delta E_{int}$ is the change of the internal energy of the molecules, $\Delta A_{int}$ is the change of the angular momentum of the collision complex, $\Delta v = v' - v$ and $\Delta j = j' - j$. 
This allows our models to classify transitions by the corresponding {\it quantum number gaps} and make predictions about cross sections for specific transitions based on information about other transitions. For example, we will illustrate that rigorous results for the $v=1, j = 0 \rightarrow v = 0, j = 2$ transition could be used to make predictions of cross sections for the $v=2, j = 0 \rightarrow v = 1, j = 2$ transition, without rigorous calculations.

We consider collisions of SiO and CO with para-H$_2$.
 The accurate and approximate results 
are obtained from 6DCC and 5DCS calculations, respectively.
Here, 6DCC stands for `six-dimensional close-coupling' and 5DCS for `five-dimensional coupled-states'. 
Cross sections at a given 
collision energy $E_c$ and wavevector $k_n$ are calculated 
using the appropriate $T$-matrix for the respective 6DCC and 5DCS formulations as follows:
\begin{align}
\sigma_{n\rightarrow n'}^{\mbox{\tiny{6DCC}}}  = 
\frac{\pi}{k_n^2 [j_1][j_2]}
\sum_{j_{12}j'_{12}ll'J} 
(2J+1)\,
\left|T^{J}_{nj_{12}l;n'j_{12}'l'} \right|^{2}
\label{6D-CC} \\
\sigma_{n\rightarrow n'}^{\mbox{\tiny{5DCS}}} =
\frac{\pi}{k_n^2 [j_1][j_2]}
\sum_{\bar{l}m_1m_2} (2\bar{l}+1)\,
\left|T^{\bar{l}m_1m_2}_{n;n'} \right|^{2},
\label{5D-CS}
\end{align}
where $[j] = 2j + 1$. 
In Eq. (\ref{6D-CC}), the $T$-matrix is diagonal with
respect to the total angular momentum quantum number $J$,
defined by the vector relations $\vec{J}=\vec{l}+\vec{j}_{12}$ 
and $\vec{j}_{12}=\vec{j}_1+\vec{j}_2$, where $\vec{l}$ is the
orbital angular momentum. 
In Eq. (\ref{5D-CS}), the $T$-matrix is independent
of $J$ and diagonal with respect to $m_1$ and $m_2$,
the projection quantum numbers of $\vec{j}_1$ and $\vec{j}_2$.
For the CS approximation, $\overline{l}\equiv J$ 
is the average value of $l$ between $|J-j_{12}|$ and $J+j_{12}$.
Details of the potential energy surfaces and scattering 
calculations for the present collision systems have been 
reported previously \cite{nature,yang1,yang3,castro,yang5}.
For the internal energy of the molecules, we use
\begin{equation}
E_{int}^{(i)}=w_e^{(i)}(v_i+1/2)-w_e^{(i)} x_e^{(i)}(v_i+1/2)^2
+B_{v_i}j_i(j_i+1)
\end{equation}
\begin{equation}
A_{int}^{(i)}=B_{v_i}(2j_i+1)
\end{equation}
\begin{equation}
B_{v_i}=B_e^{(i)}-a_e^{(i)}(v_i+1/2)
\end{equation}
with coefficients given in Ref. \cite{radzig}.

We quantify the accuracy of our results by 
root-mean-squared (RMS) relative error 
\begin{eqnarray}
\mbox{RMS relative error}
=\sqrt{\frac{1}{N}\sum_{f}^{N}\left(\frac{\sigma_{i\rightarrow f}^{exact}
-\sigma_{i\rightarrow f}^{approx}}{\sigma_{i\rightarrow f}^{exact}}
\right)^2}
\end{eqnarray}
where $\sigma^{exact}$ is the 6DCC result, $\sigma^{approx}$ 
corresponds to either the 5DCS value or the ML prediction, and $i/f$ denote the initial/final states.

\begin{figure}
\includegraphics[width=0.8\columnwidth]{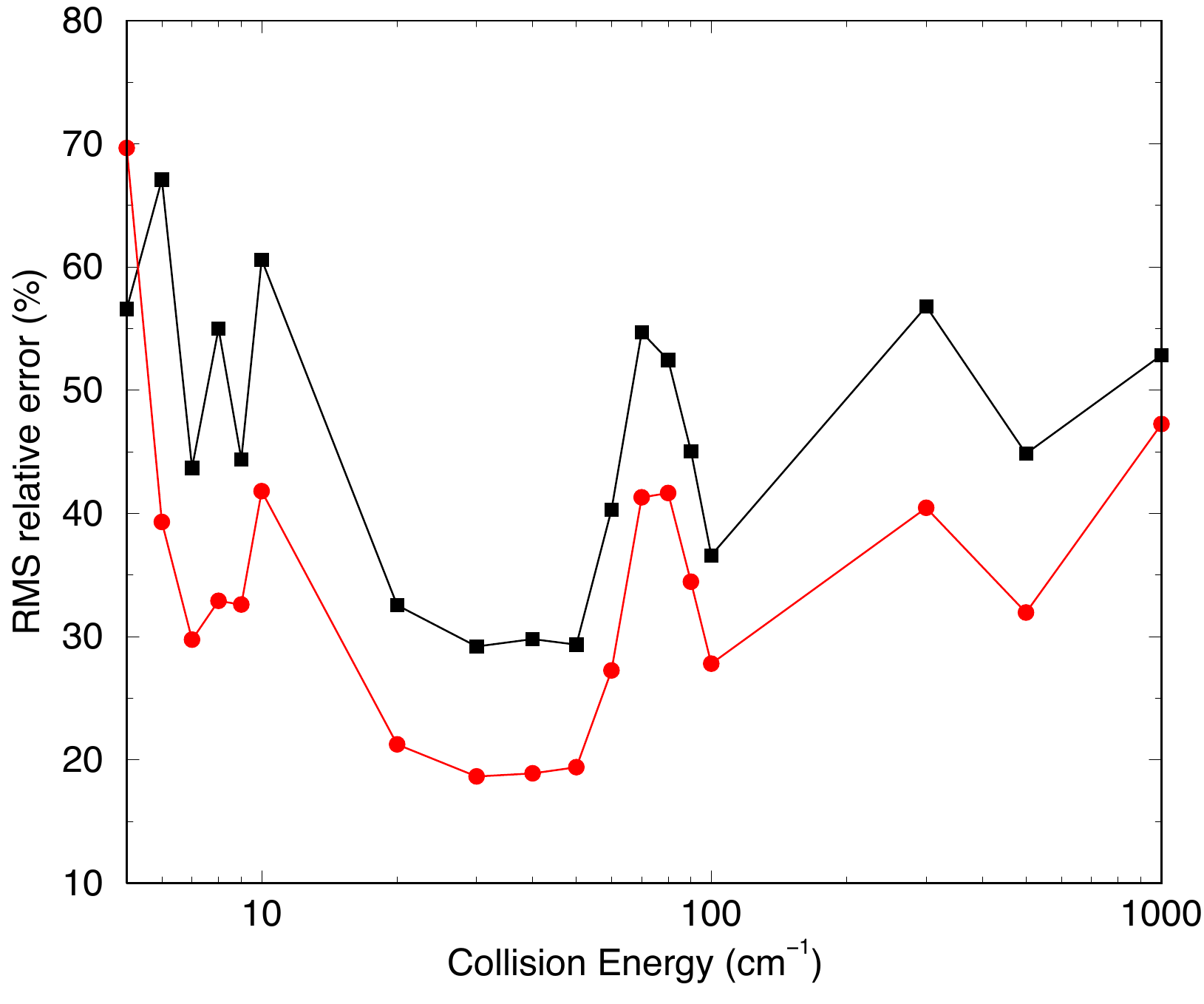}
\caption{The RMS relative error for ro-vibrational de-excitation of CO due
to collision with H$_2$: squares -- 5DCS; circles -- ML results. 
The ML model (\ref{model}) is trained using the 5DCS and 6DCC results for 21 
transitions (\ref{transition-1}) and only 5DCS for 21 transitions (\ref{transition-2}). 
No 6DCC results for transitions (\ref{transition-2}) are used to train the model (\ref{model}). 
The error depicted is for 21 transitions (\ref{transition-2}).  
}
\end{figure}





{\it Results.}
We begin by illustrating that ML can be used to improve
approximate results for ro-vibrational transitions, for which accurate 
calculations are computationally challenging. 
To do this, we consider a set of $21$ transitions
\begin{eqnarray}
\nonumber
(v_1, j_1, v_2, j_2) = (1,0,0,0) \longrightarrow (v_1', j_1', v_2', j_2') = (0, X, 0,0)
\\
\label{transition-1}
\end{eqnarray}
where $X = 0$ -- $20$. 
For each of these transitions, we calculate the 6DCC and 5DCS cross sections in the energy interval between $1$ and $1000$ cm$^{-1}$. 
Our goal is then to  predict the cross sections for a set of 21 transitions: 
\begin{eqnarray}
(2,0,0,0) \rightarrow (1, X, 0,0).
\label{transition-2}
\end{eqnarray}
without any further 6DCC calculations. We train our model (\ref{model}) by a combination of 6DCC and 5DCS cross sections for the transitions (\ref{transition-1})
and 5DCS results for the transitions (\ref{transition-2}). The model is then used to predict the cross sections for the transitions (\ref{transition-2}).
Figure 1 illustrates that the accuracy of the ML predictions is 10 to 30 \% better than of the approximate results. 

The trend observed in Figure 1 is general.  
To illustrate this, we consider transitions involving state changes of both molecules and a different system. 
We use 6DCC to compute the cross sections for 
the ro-vibrational transitions from the initial state $(v_1, j_1, v_2, j_2) = (1,4,0,0)$
for SiO($v_1, j_1$) + H$_2(v_2, j_2)$ collisions. These cross sections are then used to predict
state-resolved cross sections for the $(1,5,0,0)$ initial state 
making transitions with non-zero values of 
$\Delta v_1, \Delta j_1$ and $\Delta j_2$.
Errors are shown in Figure 2 for $\Delta v_1=-1$ and $\Delta j_2=2$.
The upper panel shows the errors calculated for all 21 transitions to the final states $(0, X = [0, 20], 0, 2)$, whereas the lower panel illustrates the reduction of the error for 10 transitions with the biggest ML improvement. 

\begin{figure}
\includegraphics[width=0.8\columnwidth]{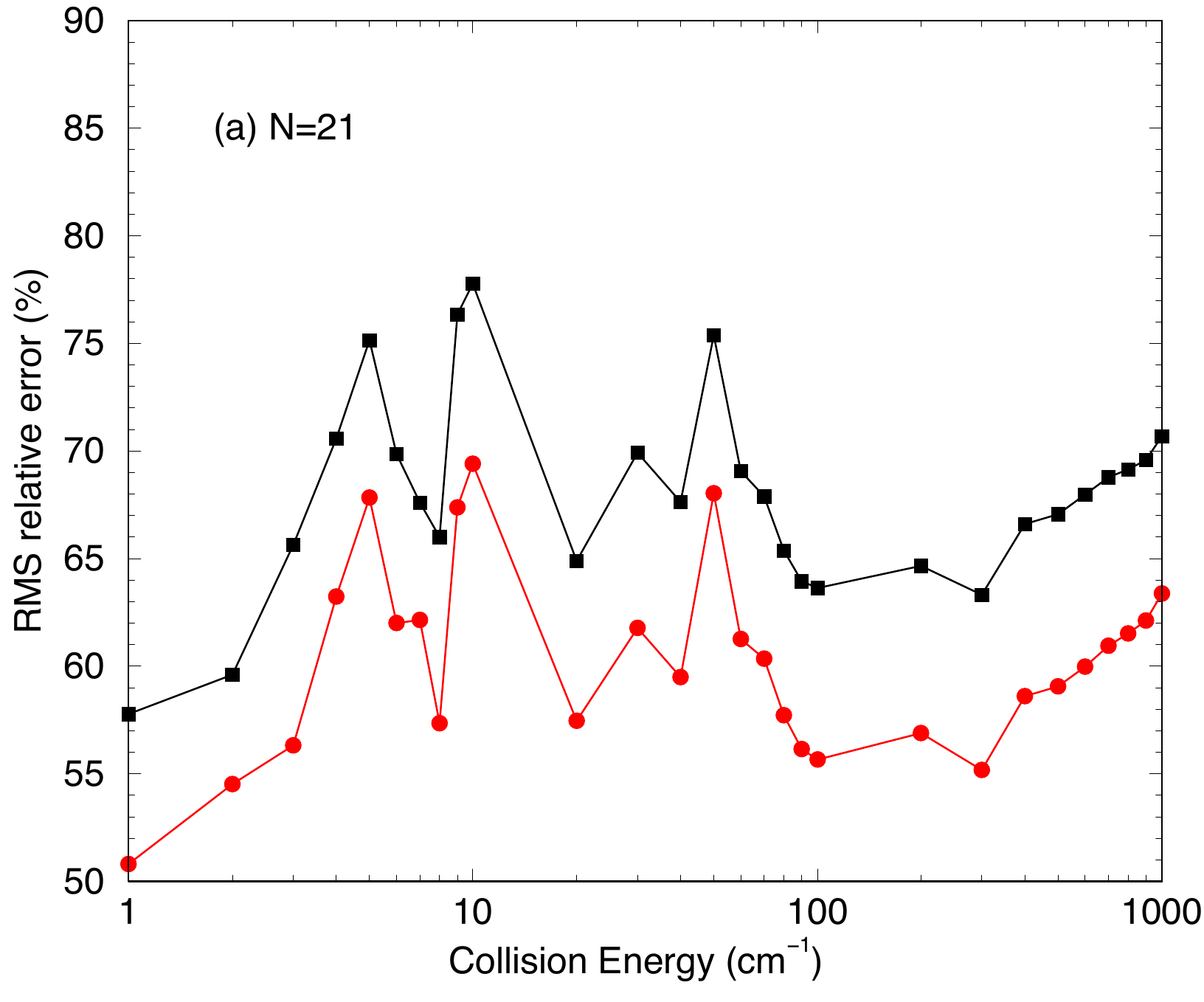} \\
\includegraphics[width=0.8\columnwidth]{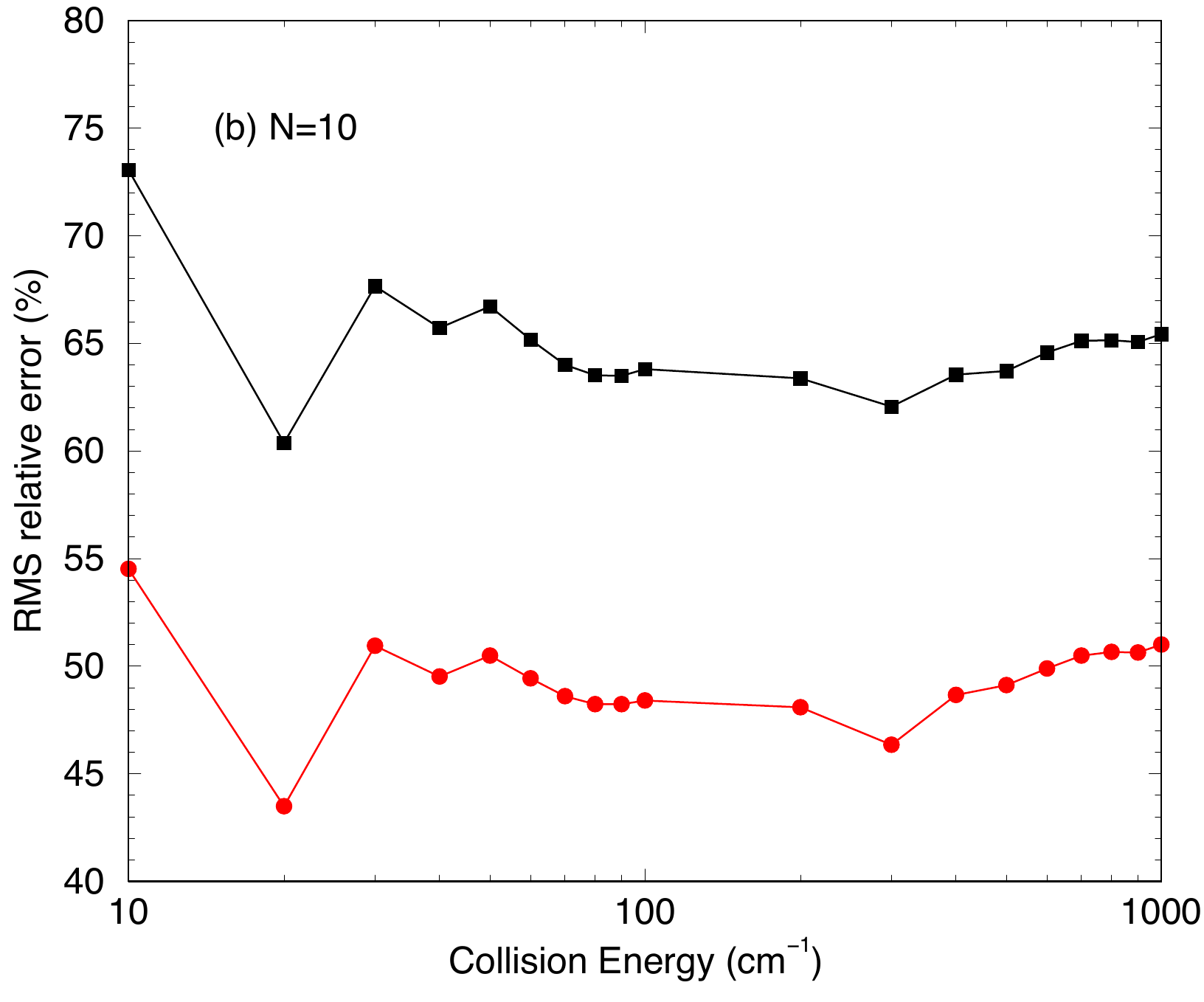} 
\caption{The RMS relative errors for ro-vibrational de-excitation of SiO
and simultaneous rotational excitation of H$_2$. 
The curves correspond to transitions from the initial state
$(v_1, j_1, v_2, j_2) = (1,5,0,0)$ to the final states $(0, X ,0,2)$, where $X=0$ -- $20$.
The black curves connecting the squares are the errors of the 5DCS results,
and the red curves connecting the circles are the errors of the ML results.
The ML models (\ref{model}) are trained by the cross sections for the transitions $(1,4,0,0) \rightarrow (0, X ,0,2)$, where $X = 0$ -- $20$. 
The lower panel illustrates the error reduction for 10 transitions with the most significant ML improvement. 
}
\end{figure}

\begin{figure}
\includegraphics[width=0.99\columnwidth]{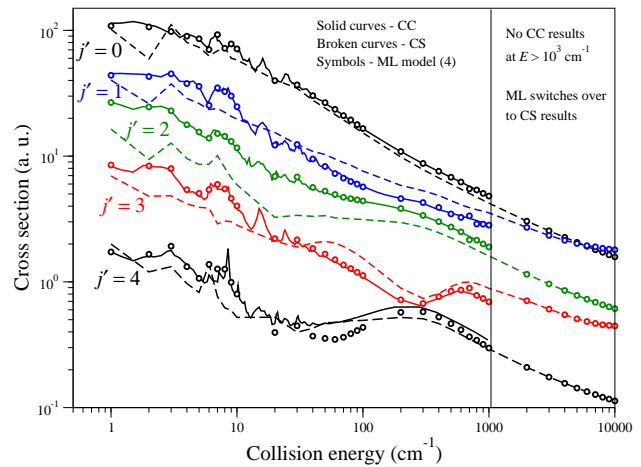}
\caption{Cross-sections for purely rotational relaxation in collisions of vibrationally ($v_1 = 1$) and rotationally ($j_1 = 5$) excited SiO molecules with H$_2$($v_2=0, j_2=0$). The final rotational state of SiO is $j'$, as indicated on the plot. 
The solid curves show the rigorous 6DCC results, the broken curves are the approximate 5DCS results and thesymbols are the ML results. 
Note that the solid curves shown in the plot are \underline{not} used for training the ML model (\ref{model}). 
The ML model (\ref{model}) is trained using the 6DCC results only for the rotationally inelastic transitions of SiO in the ground vibrational state $v_1 = 0$. There are no 6DCC results to train the model at collision energies $> 10^3$ cm$^{-1}$. As shown, at such collision energies, the model switches over to represent the CS results. 
}
\end{figure}

It should be noted that 
\begin{itemize}
\item None of the transitions considered in Figure 2 require 6DCC computations. 

\item The relative error for the $\Delta j_2=2$ transitions
is generally greater than for the corresponding $\Delta j_2=0$ transitions. 
This is due to the shortcomings of the CS approximation which decouples the angular momentum 
and assumes an average centrifugal barrier. It has been shown \cite{bohr}
that the accuracy of the CS approximation for diatom -- diatom collisions
is usually reduced when both molecules undergo a change in rotational state.

\item Despite the relatively poor starting approximation,
ML is able to significantly improve the 5DCS results over the
whole range of collision energies. 

\item  It was found to be necessary to include 
$\Delta j_2=0$ transitions in the training set in order to obtain
the $\Delta j_2=2$ results in Figure 2. The converse is not true
as the same ML results are obtained when $\Delta j_2=0$ or
$\Delta j_2=0-2$ were included in the training set.

\end{itemize}

The algorithm for improving the accuracy of the approximate calculations can also be applied to cross sections for purely rotational transitions. 
Figure 3 illustrates the accuracy of the ML model predictions of the cross sections for pure rotational relaxation of SiO molecules initially in the state ($v_1 = 1, j_1 = 5$).  
The ML model (\ref{model}) is built using the 6DCC cross sections for rotational relaxation of SiO($v_1 = 0, j_1 = 5$) molecules in the vibrationally ground state. This model is then used to correct the 5DCS results for the relaxation from the state 
($v_1 = 1, j_1 = 5$). As can be seen, the ML-corrected results are in perfect agreement with the rigorous 6DCC calculations, used here for testing purposes only. 
Note that the CC calculations for the vibrationally ground state extend only to the collision energy $10^3$ cm$^{-1}$.  Figure 3 illustrates that at energies above $10^3$ cm$^{-1}$, the ML model recovers the CS results.

The model (\ref{model}) can also be used to infer the detailed dependence of the rigorous results on the underlying Hamiltonian parameters (such as energy or total angular momentum), using a combination of a large number of fast approximate calculations and a small number of computationally demanding rigorous computations. To illustrate this, we consider chemically reactive scattering of Cl atoms with H$_2(v=1, j = 0)$ molecules. The difficulty of the quantum calculations of cross sections for atom - diatom scattering increases to a great extent as the total angular momentum of the reactive complex increases. To overcome this problem, one often uses the $J$-shifting approximation  \cite{Bowman1991, Zhang1999} to predict  the reaction cross sections for high $J$ states using the results for $J=0$. However, the $J$-shifting approximation often produces large errors, especially in the presence of quantum resonances that are inherited by the $J$-shifting approximations but may not be present in all $J$ states. This is well illustrated by Figure 4 (upper pannel), showing how the resonance features inherited from the $J=0$ dynamics affect the $J$-shifting predictions of the cross sections for $J=20$. 

An alternative to the $J$-shifting calculations can be our model (\ref{model}). The ML results for the upper panel are obtained using 51 rigorous calculations for the energy dependence of the cross section at $J=0$ and 8 rigorous calculations for $J=20$.  In this case, the $J=0$ calculations play the role of the approximate results and $8$ of the $J=20$ cross sections are used to correct the energy dependence of the cross sections, completely removing the resonance structure. To illustrate the generality of this approach, we repeat the calculation for $J=60$ and $J=65$. The $J$-shifting results for such high angular momenta are completely unphysical, while the ML model produces an accurate energy dependence of the cross sections.

In summary, we have illustrated a general, {\it system and approximation-independent}, approach to improve the accuracy of quantum dynamics approximations. The main idea is based on Eq. (\ref{model}) that uses two independent GPs: one  to infer the general dependence of the dynamical results on the underlying Hamiltonian parameters {and} one to infer the difference between the approximate and rigorous results. The most straightforward application of this approach is to save CPU time by 
interpolating a small number of rigorous results with a large number of ML-corrected approximate calculations. More importantly, the present work demonstrates, that if the input space variables are designed as in Eq. (\ref{xset}), the model (\ref{model}) can generalize the correlations between the approximate and rigorous results to a range of different transitions. Thus, a model (\ref{model}) trained by  a combination of approximate and rigorous results for a certain range of transitions can make accurate predictions for different transitions without rigorous calculations. This opens the possibility of improving the accuracy of approximate calculations for quantum transitions that are out of reach of rigorous CC calculations. 

\begin{figure}
\includegraphics[width=0.8\columnwidth]{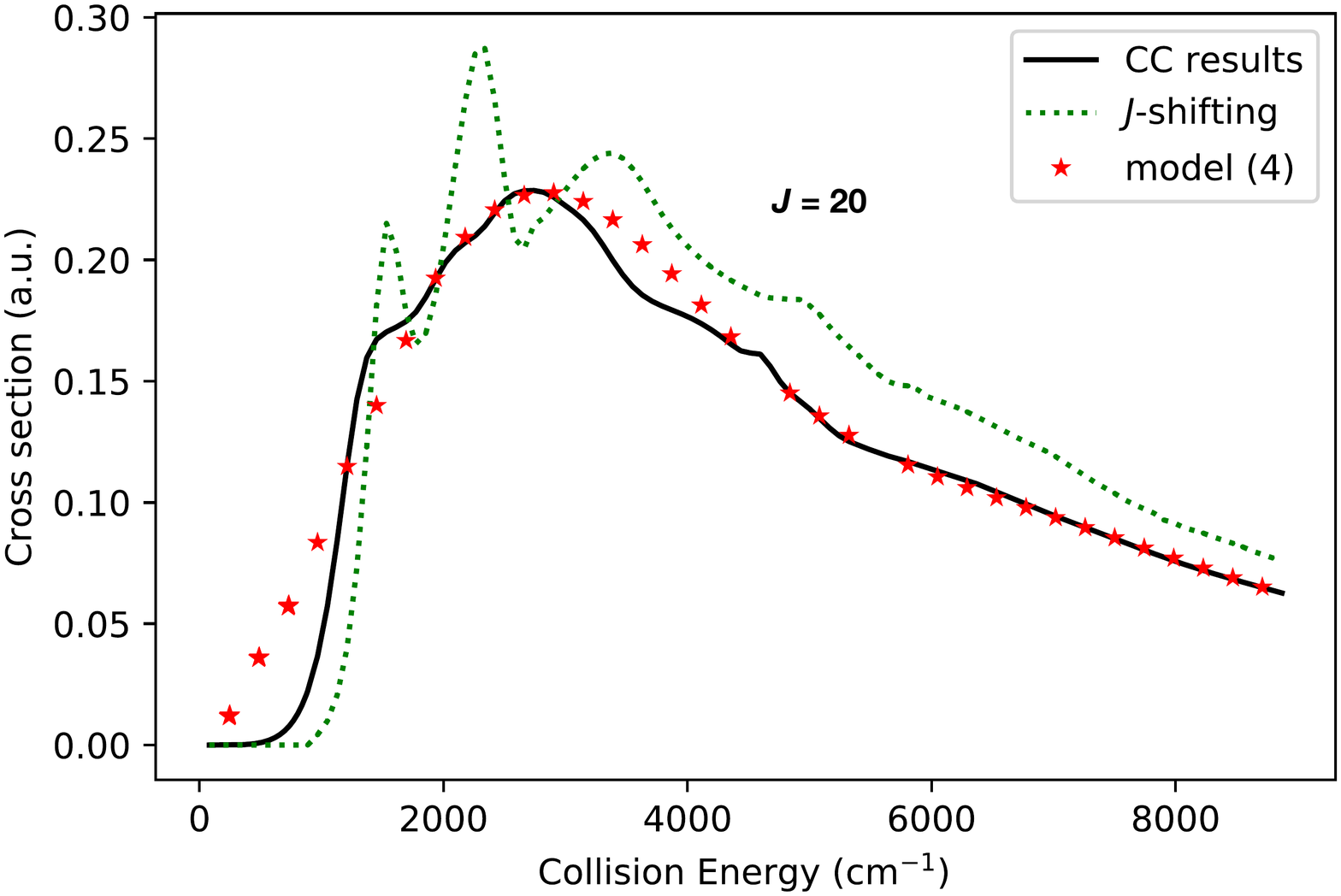} \\
\includegraphics[width=0.8\columnwidth]{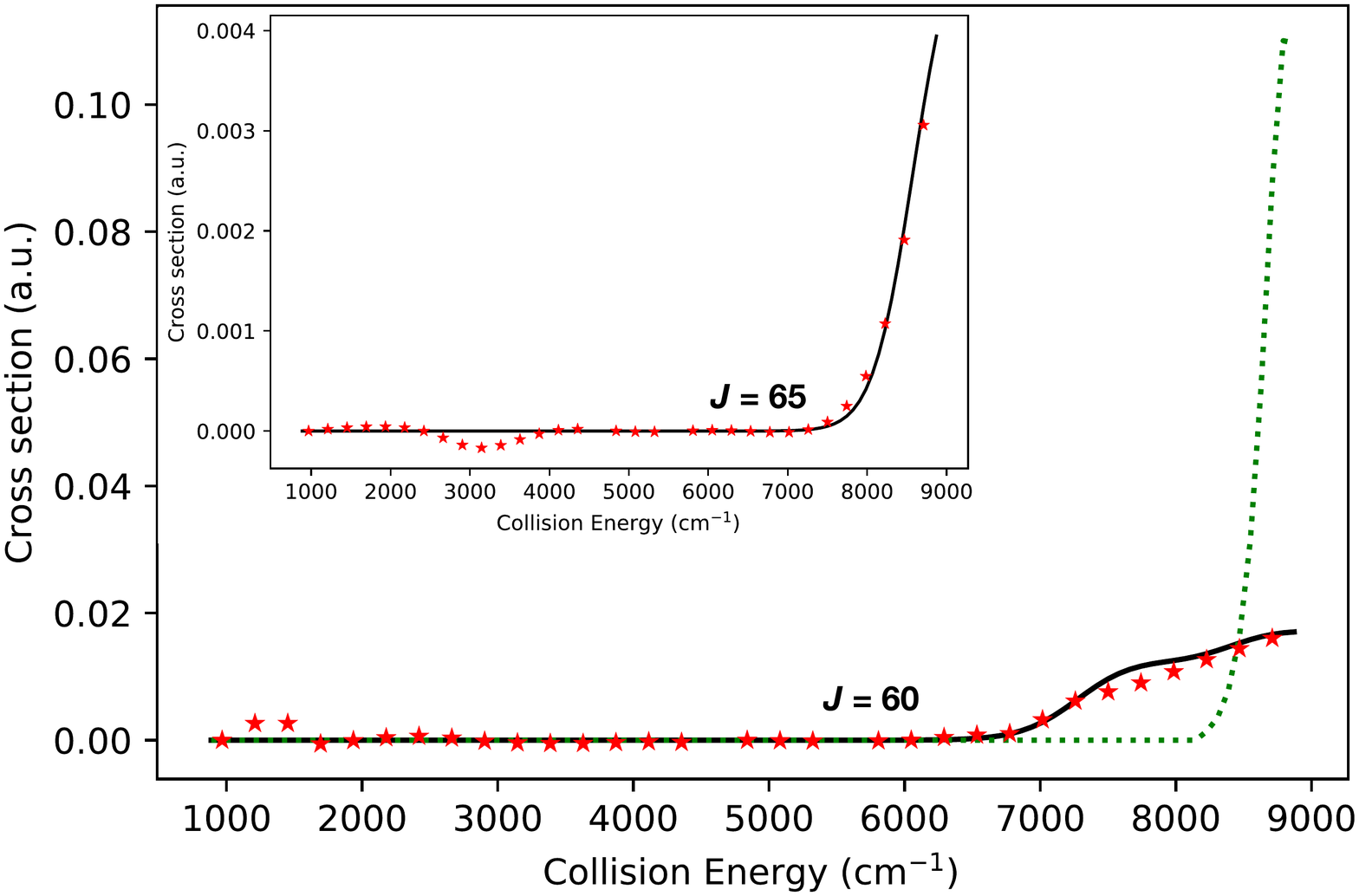} 
\caption{Energy dependence of the total cross section for the chemical reaction Cl + H$_2(v=1, j = 0)$ $\rightarrow$ HCl + H: the solid lines -- rigorous close coupling calculations, the dotted curves  -- the $J$-shifting results, the symbols -- the predictions using the ML model (\ref{model}). The ML model is trained by a combination of 51 cross sections for $J = 0$ and $8$ cross sections for the corresponding high value of $J$.   
}
\end{figure}

\begin{acknowledgments}
This work was supported at Penn State 
by NSF Grant No. PHY-1806180, at UGA by NASA grant NNX16AF09G, at UNLV by NSF grant No. 1806334, and by NSERC of Canada.
\end{acknowledgments}


\begin{thebibliography}{99}

\bibitem{alex}
A. M. Arthurs A. M. Arthurs, 
The theory of scattering by a rigid rotator, 
Proc. Royal. Soc. London {\bf 256}, 540 (1960). 

\bibitem{mona}
M. Berciu,
Few-Particle Green's Functions for Strongly Correlated Systems on Infinite Lattices,
 Phys. Rev. Lett. {\bf 107}, 246403 (2011).

\bibitem{jeremy}
J. M. Hutson,
Coupled channel methods for solving the bound-state Schr\"{o}dinger equation,
Comp. Phys. Comm. {\bf 84}, 1 (1994). 

\bibitem{bmc}
M. C. Kennedy and A. O`Hagan, Bayesian calibration of computer models, J. Royal Stat. Soc. B 63,
425 (2001). 


\bibitem{lique}
E. Roueff and F. Lique, 
Molecular excitation in the interstellar medium: recent advances in collisional, radiative, and chemical processes, 
Chem. Rev. {\bf 113}, 8906 (2019). 

\bibitem{Perreault2017} W. E. Perreault, N. Mukherjee, and R. N. Zare, Quantum control of molecular collisions at 1 Kelvin, Science {\bf 358}, 356 (2017).

\bibitem{Krems2008} R. V. Krems, Cold Controlled Chemistry, Phys. Chem. Chem. Phys. {\bf 10}, 4079 (2008).

\bibitem{Bala2016} N. Balakrishnan, Perspective: Ultracold molecules and the dawn of cold controlled chemistry, J. Chem. Phys. {\bf 145}, 150901 (2016).

\bibitem{Jambrina2019} P. G. Jambrina, J. F. E. Croft, H. Guo, M. Brouard, N. Balakrishnan, and F. J. Aoiz, Stereodynamical control of a quantum scattering resonance
in cold molecular collisions, Phys. Rev. Lett. {\bf 123}, 043401 (2019).

\bibitem{my-book}
R. V. Krems, Molecules in electromagnetic elds: from ultracold physics to controlled chemistry (Wiley, New York, 2018).


\bibitem{Bowman1991} J. M. Bowman, Reduced dimensionality theory of quantum reactive scattering, J. Phys. Chem. {\bf 95}, 4960 (1991).

\bibitem{Zhang1999} D. H. Zhang and J. Z. H. Zhang, Uniform $J$-shifting approach for calculating reaction rate constant, J. Chem. Phys. {\bf 110}, 7622 (1999).



\bibitem{qcp-1}
G. Carleo and M. Troyer,
{Solving the quantum many-body problem with artificial neural networks},
{Science} {\bf 355}, 602 (2017).

\bibitem{qcp-2}
K. Ch'ng, J. Carrasquilla, R. G. Melko, and E. Khatami,
{Machine Learning Phases of Strongly Correlated Fermions},
{Phys. Rev. X} {\bf 7}, 031038 (2017).

\bibitem{qcp-3}
H. Saito,
{Solving the Bose-Hubbard model with machine learning},
{J. Phys. Soc. Jpn} {\bf 86}, 093001 (2017).

\bibitem{qc-1}
Y. Nomura, A. S. Darmawan, Y. Yamaji, and M. Imada,
{Restricted Boltzmann machine learning for solving strongly correlated quantum systems}, {Phys. Rev. B} {\bf 96}, 205152 (2017).

\bibitem{qc-2}
O. A. von Lilienfeld,
{Quantum Machine Learning in Chemical Compound Space},
{Angew. Chem. Int. Ed.} {\bf 57}, 4164 (2018).

\bibitem{qc-3}
J. Han, L. Zhang, and E. Weinan,
{Solving many-electron Schr\"{o}dinger equation using deep neural networks},
{J Comp. Phys.} {\bf 399}, 108929 (2019).

\bibitem{qc-4}
K.T. Sch\"{u}tt, M. Gastegger, A. Tkatchenko, K. R. M\"{u}ller, and R. J. Maurer,
{Unifying machine learning and quantum chemistry with a deep neural network for molecular wavefunctions},
{Nat Commun} {\bf 10}, 5024 (2019).

\bibitem{qc-5}
K. Choo, A. Mezzacapo, G. Carleo,
{Fermionic neural-network states for ab-initio electronic structure},
arXiv:1909.12852

\bibitem{qc-6}
D. Luo and B. K. Clark,
{Backflow Transformations via Neural Networks for Quantum Many-Body Wave Functions},
{Phys. Rev. Lett.} Ê{\bf 122}, 226401 (2019).

\bibitem{qc-7}
J. Hermann, Z. Sch\"{a}tzle and F. No\'{e},
{Deep neural network solution of the electronic Schr\"{o}dinger equation},
arXiv:1909.08423


\bibitem{md-1}
F. No\'{e}, S Olsson, J. K\"{o}hler, H. Wu,
{Boltzmann generators: Sampling equilibrium states of many-body systems with deep learning},
{Science} {\bf 365}, 6457 (2019).

\bibitem{md-2}
 D. Koner,  O. T. Unke,  K. Boe,  R. J. Bemish, and  M. Meuwly, 
Exhaustive state-to-state cross sections for reactive molecular collisions from importance sampling simulation and a neural network representation,
 \jcp{150}{211101}{2019}.

\bibitem{gp-book}
C. E. Rasmussen, and C. K. I. Williams, 
{\it Gaussian Processes for Machine Learning} (The MIT Press, Cambridge, 2006).



\bibitem{extrapolation-1}
D. K.  Duvenaud, H. Nickisch, and C. E. Rasmussen,
{Additive Gaussian Processes}, 
{\it Adv. Neur. Inf. Proc. Sys.} {\bf 24}, 226 (2011).

\bibitem{extrapolation-2}
D. K. Duvenaud, J. Lloyd, R. Grosse, J. B. Tenenbaum, and Z. Ghahramani,
{Structure Discovery in Nonparametric Regression through Compositional Kernel Search},
{\it Proceedings of the 30th International Conference on Machine Learning Research} {\bf 28}, 1166 (2013).


\bibitem{extrapolation-3}
R.Vargas-Hernandez, J. Sous, M. Berciu, and R. V. Krems, 
{Extrapolating quantum observables with machine learning: Inferring multiple phase transitions from properties of a single phase}, 
\prl{121}{255702}{2018}.


\bibitem{NNs-vs-GPs}
A. Kamath, R. A. Vargas-Hernandez, R. V. Krems, T. Carrington Jr., and S. Manzhos, {Neural networks vs Gaussian process regression for representing potential energy surfaces: A comparative study of fit quality and vibrational spectrum accuracy},
\jcp{148}{241702}{2018}.



\bibitem{rodrigo-bo}
R.Vargas-Hernandez, Y. Guan, D. H. Zhang, and R. V. Krems, {Bayesian optimization for the inverse scattering problem in quantum reaction dynamics}, 
{\it New J. Phys.} (Fast Track Communication) {\bf 21}, 022001 (2019). 


\bibitem{bml} R. V. Krems, 
Bayesian machine learing for quantum molecular dynamics,
Phys. Chem. Chem. Phys. {\bf 21}, 13392 (2019).


\bibitem{krems2} J. Cui and R. V. Krems, 
Gaussian process model for collision dynamics of complex molecules,
Phys. Rev. Lett. {\bf 115}, 073202 (2015).



\bibitem{qc-bmc}
A. E. Wiens, A. V. Copan, H. F. Schaefer, {Multi-Fidelity Gaussian Process Modeling for Chemical Energy Surfaces}, 
{\it Chem. Phys. Lett.} X, in press (2019). 



\bibitem{nature} B. H. Yang, P. Zhang, X. Wang, P. C. Stancil, J. M. Bowman, 
N. Balakrishnan, and R. C. Forrey,  
Quantum dynamics of CO-H$_2$ in full dimensionality,
Nat. Comm. {\bf 6}, 6629 (2015).

\bibitem{yang1} B. H. Yang, N. Balakrishnan, P. Zhang, X. Wang, J. M. Bowman,
R. C. Forrey, and P. C. Stancil,
Full-dimensional quantum dynamics of CO in collision with H$_2$, 
J. Chem. Phys. {\bf 145}, 034308 (2016).


\bibitem{yang3} B. H. Yang, P. Zhang, C. Qu, H. Wang, P. C. Stancil,
J. M. Bowman, N. Balakrishnan, B. M. McLaughlin, and R. C. Forrey,
Full-Dimensional Quantum Dynamics of SiO in Collision with H$_2$, 
J. Phys. Chem. A {\bf 122}, 1511 (2018).











\bibitem{castro} C. Castro, K. Doan, M. Klemka, R. C. Forrey, B. H. Yang, P. C. Stancil, and N. Balakrishnan, 
Inelastic cross sections and rate coefficients for collisions between CO and H$_2$, 
Mol. Astrophys. {\bf 6}, 47 (2017). 

\bibitem{yang5} B. H. Yang, P. C. Stancil, N. Balakrishnan, and R. C. Forrey, 
Rotational quenching of CO due to H$_2$ collisions, 
Astroph. J. {\bf 718}, 1062 (2010). 

\bibitem{radzig} A. A. Radzig and B. M. Smirnov,
Reference Data on Atoms, Molecules, and Ions,
Springer Series in Chemical Physics, vol. 31 
(edited by J. P. Toennies).


\bibitem{bohr} A. Bohr, S. Paolini, R. C. Forrey, N. Balakrishnan, and 
P. C. Stancil,  
A full-dimensional quantum dynamical study of H$_2$+H$_2$ collisions: 
coupled-states versus close-coupling formulation,  
J. Chem. Phys. {\bf 140}, 064308 (2014).


\end{thebibliography}
\end{document}